\documentclass[preprint,prl,aps,singlecolumn,superscriptaddress]{revtex4-2}
\usepackage{bm}
\usepackage[colorlinks=true,linkcolor=blue,citecolor=blue]{hyperref}
\usepackage{times}
\usepackage{amsmath}
\usepackage{amssymb}
\usepackage{physics}
\usepackage{amsthm}
\usepackage{amsfonts}
\usepackage{enumerate}
\usepackage{latexsym}
\usepackage{ifpdf}
\usepackage{cancel}
\usepackage{natbib}
\usepackage{psfrag}
\usepackage[utf8]{inputenc}
\usepackage[T1]{fontenc}
\usepackage{mathtools}
\usepackage[thinc]{esdiff}
\newcommand{\beq}{\begin{equation}}
	\newcommand{\eeq}{\end{equation}}
\usepackage{graphicx}
\usepackage{makeidx}
\usepackage{color}
\usepackage{array}
\usepackage{multirow}
\usepackage{siunitx}
\usepackage{physics}

\begin{document}
\title{ Beyond being free: glassy dynamics of SrTiO$_3$-based two-dimensional electron gas}
\author {Jyotirmay Maity}
\altaffiliation{Contributed equally}
\affiliation{Department of Physics, Indian Institute of Science, Bengaluru 560012, India}

\author {Shashank Kumar Ojha}
\email{shashank@iisc.ac.in}
\altaffiliation{Contributed equally}
\affiliation{Department of Physics, Indian Institute of Science, Bengaluru 560012, India}

 \author {Prithwijit Mandal}
\affiliation{Department of Physics, Indian Institute of Science, Bengaluru 560012, India}
 \author {Manav Beniwal}
\affiliation{Department of Physics, Indian Institute of Science, Bengaluru 560012, India}
 \author {Nandana Bhattacharya}
\affiliation{Department of Physics, Indian Institute of Science, Bengaluru 560012, India}
\author {Andrei Gloskovskii}
	\affiliation{Deutsches Elektronen-Synchrotron DESY, 22607 Hamburg, Germany}
	\author {Christoph Schlueter}
	\affiliation{Deutsches Elektronen-Synchrotron DESY, 22607 Hamburg, Germany}
\author {Srimanta Middey}
\email{smiddey@iisc.ac.in}
\affiliation{Department of Physics, Indian Institute of Science, Bengaluru 560012, India}

\textbf{	\begin{abstract}
Electron glasses offer a convenient laboratory platform to study glassy dynamics. 
Traditionally, the interplay between long-range Coulomb interactions and disorder is deemed instrumental in stabilizing the electron glass phase. Existing experimental studies on electron glass have focused on doped semiconductors, strongly correlated systems, granular systems, etc., all of which are far from the well-delocalized limit. In this work we expand the study of electron glasses to a well-known quantum paraelectric SrTiO$_3$ (STO) and unveil a new scenario: how naturally occurring ferroelastic twin walls of STO could result in glassy electrons, even in a metallic state. We show that the emergent two-dimensional electron gas at the $\gamma$-Al$_2$O$_3$/STO interface exhibits long-lasting temporal relaxations in resistance and memory effects at low temperatures, which are hallmarks of glassiness. We also demonstrate that the glass-like relaxations could be further tuned by application of an electric field.
This implies that the observed glassy dynamics is connected with the development of polarity near the structural twin walls of STO and the complex interactions among them, arising from the coupling between ferroelastic and ferroelectric orders.
The observation of this glassy metal phase not only extends the concept of electron glasses to metallic systems with multiple order parameters but also contributes to the growing understanding of the fascinating and diverse physical phenomena that emerge near the quantum critical point.
\end{abstract} }

 \maketitle
 
	{\color{magenta}{\bf{Introduction:}}}
Understanding the nature of low-energy electronic excitations in strongly disordered insulators is a key challenge in condensed matter physics~\cite{Anderson:1958p1492,Mott2004metal,RevModPhys.57.287}.
Disorder inherently promotes local density fluctuation, while Coulomb repulsion favors a more uniform distribution.  This interplay leads to a multitude of metastable states with comparable energies. The resulting phase exhibits slow relaxations, aging, and memory effects, etc and known as electron glass~\cite{Davies:1982p758,Amir:2011p235,Dobrosavljevic:2012p,Pollak:2013p}.
Following the initial observation  in granular system~\cite{Adkins:1984p4633}, the electron glass phase has been identified in diverse material platforms, including complex oxides (e.g. high-$Tc$ cuprates, manganites, ruthenates, vanadates etc.)~\cite{Dagotto:2005p257,Raiifmmode:2008p177004,Samizadeh:2022p596}, MOSFET like (metal- oxide-semiconductor field-effect transistor) structure~\cite{Vaknin:2000p3402,Jaroszyifmmode:2006p037403}, three-dimensional doped semiconductors (Si:P)~\cite{Kar:2003p216603}. All of these systems are away from good metallic regime as well-delocalized electrons are expected to exhibit a single, well-defined ground state, precluding glassiness. In this work, we report the observation of glassy electron dynamics in a completely new platform: SrTiO$_3$, a well-known quantum parelectric, transformed into a metal via electron doping.

\begin{figure*}
		\centering{
			{~}\hspace*{0cm}
			\includegraphics[scale=.38]{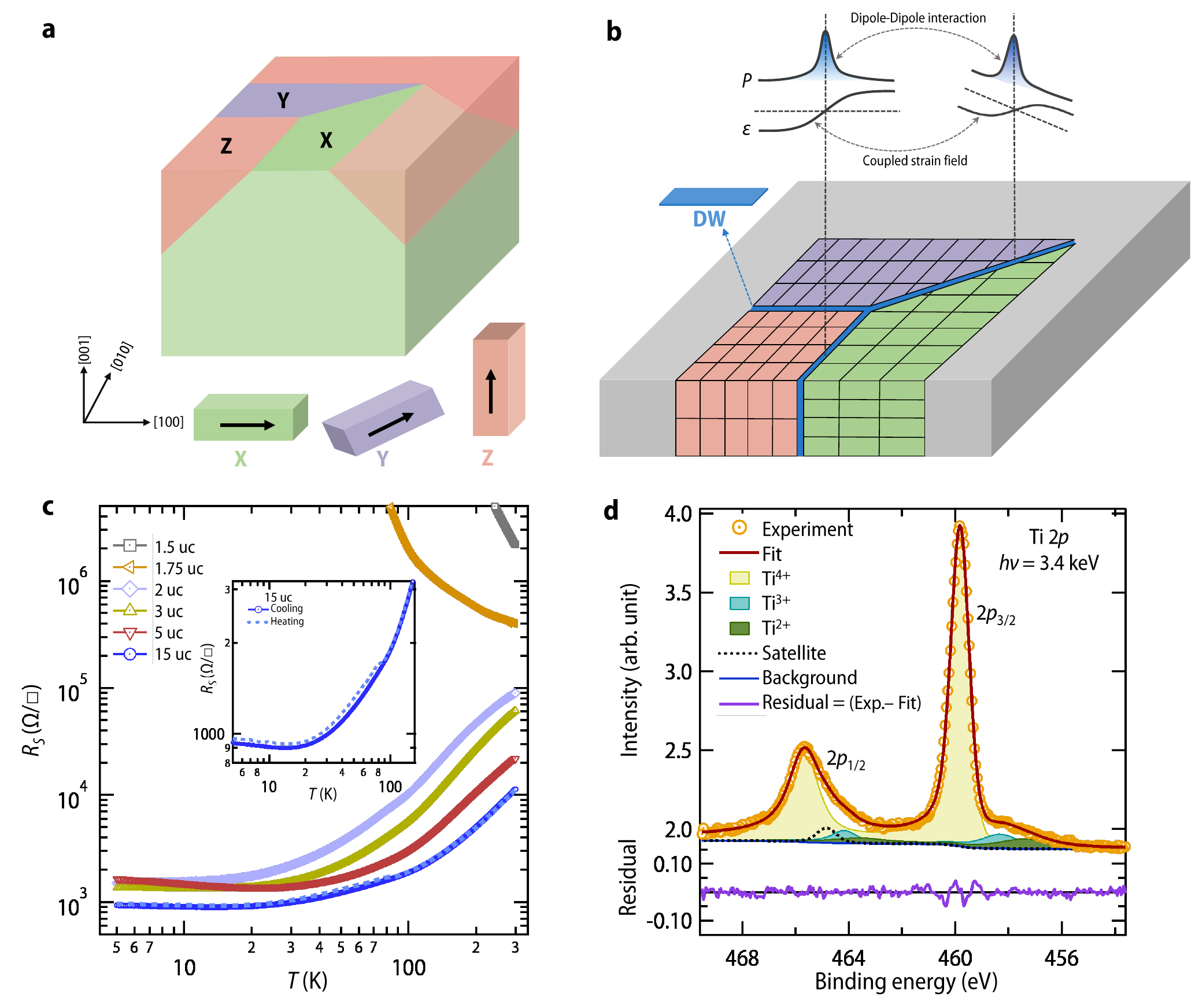}
			\caption{\textbf{Tetragonal domains in STO and electronic structure of GAO/STO heterostructures} \textbf{a.} A schematic showing structural twin domains of STO below 105 K. Three types of domains with $c$ axis along [100], [010], and [001] have been marked with X (green), Y (purple), and Z (red) respectively~\cite{Cowley:1964pA981}.  \textbf{b.} Schematic to depict coexistence  of strain field ($\epsilon$) and polarization ($P$) within the domain wall. For the sake of simplicity, we have only shown the component which is perpendicular to the domain wall. 
   A spontaneous strain field arises due to inherent ferroelasticity, which also varies in sign across the domain wall~\cite{Nataf:2020p634}. Further, the maximum value of polarisation appears at the domain wall. Within a domain wall, both polarization and strain field are coupled. Additionally, one domain wall can also interact with the neighbouring one via either dipole-dipole interaction or through coupled strain fields~\cite{Pesquera:2018p235701,Kustov:2020p016801}. \textbf{c.} Temperature-dependent sheet resistance ($R_S$) in the cooling cycle (ramp rate = 3 K/min) for GAO/STO heterostructures with different thicknesses, demonstrating a metal to insulator transition with lowering the film thickness. { Inset shows the hysteresis in $R_s$ vs. $T$ plot between heating and cooling run for the 15 uc GAO/STO sample.} \textbf{d.} Upper panel shows measured Ti 2$p$ core level spectrum of a metallic GAO/STO sample using HAXPES. Fitting of the spectra reveals the presence of Ti$^{3+}$ and Ti$^{2+}$ in the sample, in addition to Ti$^{4+}$. A charge transfer shake-up satellite peak \cite{Sen:1976p560} was also included in the fitting. The bottom panel shows the residual of the fitting, testifying the excellent matching with the measured spectrum.}  \label{fig:1}}
	\end{figure*}

In pristine form, STO is a cubic perovskite at room temperature with a band gap of 3.25 eV~\cite{Cardona:1965pA651}. It undergoes a cubic to tetragonal structural phase transition at  $T_\textnormal{C}$ $\sim$ 105 K driven by a transverse acoustic phonon mode at $R$ point of the Brillouin zone~\cite{Cowley:1964pA981}.  This transition, also known as antiferrodistortive (AFD) transition, results in a dense network of ferroelastic twin domains with distinct walls separating each group of domains (see Fig. \ref{fig:1}a). The transverse optical phonon mode at $\Gamma$-point softens upon lowering the temperature and saturates around 2 meV below 35 K, leading to the well known quantum paraelectic phase~\cite{Muller:1979p3593,Rowley:2014p367}. Most importantly, several studies have further corroborated the emergence of finite polarity at the AFD domain wall (DW) at 80 K, which enhances further below 40 K, though the individual domains remain nonpolar~\cite{Petzelt:2001p184111,Zubko:2007p1667601,Scott:2012p187601,Salje:2013p247603}. This is attributed to the coupling between the ferroelastic and ferroelectric order parameters (Fig. \ref{fig:1}b)~\cite{Morozovska:2012p094107}. 
Furthermore, the complex interactions among these polar DWs, in presence of quantum fluctuation over mesoscopic length-scale (due to hybridized acoustic-optical phonons), leads to the emergence of various collective phases, including quantum domain glass and quantum domain solid, as the temperature is varied~\cite{Pesquera:2018p235701,Kustov:2020p016801,Fauque:2022pL140301}.

Metallic STO, obtained through either chemical doping or by heterostructuring with another insulator, hosts various quantum phenomena including superconductivity, ferromagnetism, Rashba spin orbit coupling, polar metal etc. and has become a research theme on its own~\cite{Frederikse:1964pA442,Ohtomo:2004p423,Reyren:2007p1196,Caviglia:2010p126803,Li:2011p762,Bert:2011p767,Rischau:2017p643,Collignon:2019p031218,Pai:2018p036503,Cao:2018p1547,Brehin:2023p823}. While the electrical dipoles are expected to be screened by the mobile carriers, several  observations including  enhanced conductivity along the DWs, stress tunable conductivity, zero-field transverse resistance, charge trapping within domain walls etc. firmly 
establish that the polar DWs have a significant influence on electrical conduction in  STO~\cite{Kalisky:2013p1091,Harsan:2016p257601,Frenkel:2017p1203,Christensen:2019p269,Krantz:2021p036801,ojha:2021p054008}. 
Despite these investigations into the static influence of DWs on electrical transport, there is no report of dynamical impact of DWs on conduction electrons. In this work, we explore whether the glassy DWs in the quantum domain glass phase could impart glassiness to the conduction electrons in metallic STO. Delving in this direction would not only advance our understanding of electron glasses beyond traditional models but it would also be extremely crucial for understanding  electronic transport in polar metals and polar superconductors, a topic of significant current interest~\cite{Bhowal:2023p53,Rischau:2017p643,Wang:2019p61,Volkov:2022p4599,Yu:2022p63}.

This work presents the first observation of glassy dynamics in a STO-based two-dimensional electron gas (2DEG). For this, we have focused on the 2DEG obtained by growing $\gamma$-Al$_2$O$_3$  (GAO) film on STO (001) substrate due to its exceptional sensitivity to DWs in electrical transport~\cite{Christensen:2019p269}. 
We found that this metallic interface exhibits a gradual relaxation of resistance at fixed temperatures, similar to conventional electron glasses. Interestingly, the relaxation magnitude increases below 80 K, the same temperature where the polarity emerges within the DWs of pristine STO~\cite{Scott:2012p187601}. The most significant relative change in resistance relaxation is observed at 22 K, and the relaxation disappears below 12 K. Notably, these resistance relaxations depend on the cooling rate and exhibit a memory effect as well, which are prime signatures of glassy dynamics~\cite{Dobrosavljevic:2012p,Pollak:2013p,Amir:2011p235}. Remarkably, all these observed temperature scales align with the domain glass and domain solid phases, previously reported in pristine STO~\cite{Kustov:2020p016801}, emphasizing the dynamical impact of polar DWs on electronic transport.

%{\color{magenta}{\bf{Results:}}} 

{\color{magenta}{\bf{Sample growth and characterization:}}} 
 GAO is a cubic spinel with a lattice constant
of 7.911 \AA, almost double of the lattice constant of cubic STO (3.905 \AA). Epitaxial thin film of GAO with excellent morphological quality can be grown  on STO (001) substrate due to the close match between the
oxygen sublattices of the two compounds~\cite{Chen2013,ojha:2021p054008}. For the present investigation, a series of GAO films with varying thickness on STO were grown by pulsed laser deposition technique (details of growth are in Methods section). For electrical resistance measurements, the Hall bar structure was made~\cite{Ojha:2023p126}, and the Ohmic contacts were established by using an ultrasonic wire bonder (measurement details are in the Methods section). 
Fig. \ref{fig:1}c shows the temperature-dependent sheet resistance ($R_S$) plot for the heterostructures showing a metal-insulator transition as a function of GAO film thickness. As evident, heterostructures with film thickness greater than 1.75 uc (uc=unit cell) display metallic behavior. To probe the origin and location of the conducting electrons, we carried out hard X-ray photoelectron
spectroscopy (HAXPES) on Ti 2$p_{3/2}$ and 2$p_{1/2}$ core levels of an 8.5 uc (6.8 nm) GAO/STO sample using   3.4 keV photon at the P22 beamline of PETRA III, DESY (other details are in Methods section). Apart from the strong features of Ti$^{4+}$, small shoulders are observed at lower binding energies (Fig. \ref{fig:1}d), which can be captured considering the presence of additional Ti$^{3+}$ and Ti$^{2+}$. This confirms that the conduction electrons are located within STO. The origin of the 2DEG can be further assigned to the scavenging of oxygen within the interfacial layers of STO~\cite{Chen2013,Cao:2016p1}. 
 
We further note that a noticeable difference in resistance between heating and cooling runs is observed at lower temperatures, which disappears around 90 K for all of these metallic samples (Inset of Fig. \ref{fig:1}c). Similar hysteresis below 90 K was reported in LaAlO$_3$/SrTiO$_3$ 2DEG as well~\cite{Goble:2017p44361} and was ascribed to the development of polarity near the DWs of STO~\cite{Scott:2012p187601,Salje:2013p247603}. As our investigation centres on glassy dynamics within a well-delocalized metallic regime, the remainder of this paper will focus on the 15 uc GAO/STO thin film unless specified otherwise.

	\begin{figure*}[ht]
		\centering{
			{~}\hspace*{-0.45cm}
			\includegraphics[scale=.65]{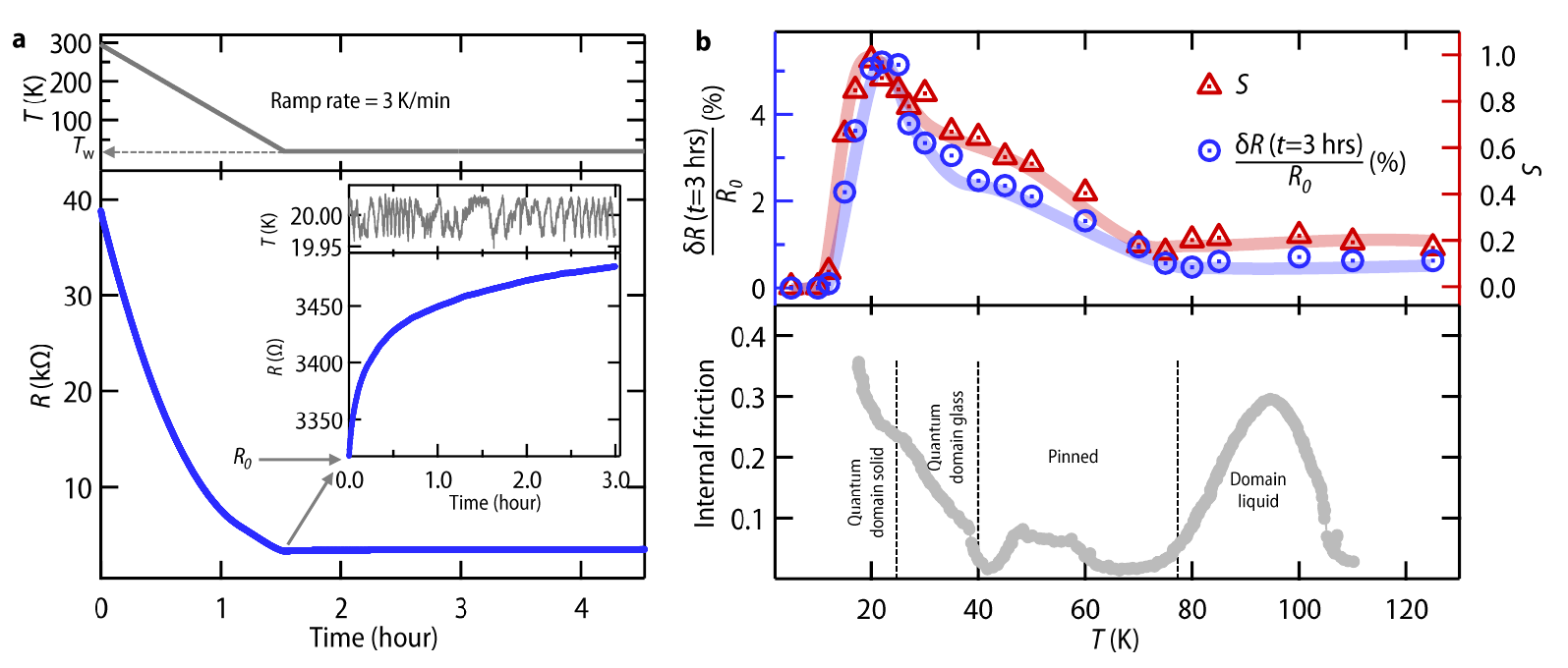}
			\caption{\textbf{Demonstration of resistance relaxations as a function of time at various fixed temperatures}  \textbf{a.} The upper panel shows the temperature ramping protocol for measuring the dynamical impact of DWs. $T_w$ is the final temperature at which the temporal evolution of resistance is measured. The lower panel shows the temporal evolution of resistance. The inset in the lower panel shows a representative example of the temporal evolution of resistance at a fixed temperature of 20 K. $R_0$ is the bare resistance just at the start of the relaxation. For further consideration, we redefine the $t$=0 at the start of the relaxation. \textbf{b.} The upper panel shows the relative percentage change of resistance at the end of three hours $\frac{\delta R \textnormal{(t=3 hrs)}}{R_0}(\%)$ and relaxation rate ($S$ =$ \frac{d\left[\frac{\delta R(t)}{R_0}\right]}{d\ln(t)}$) at several fixed temperatures. $\delta$$R(t)$ is defined as $R$($t$)-$R_0$.   The lower panel shows the temperature-dependent ultrasonic internal friction of bulk STO. This panel has been reproduced with permission from~\cite{Kustov:2020p016801}. }\label{fig:2}} 
            \end{figure*}

\begin{figure*}[htp]
		\centering{
			{~}\hspace*{-0.2cm}
			\includegraphics[scale=.7 ]{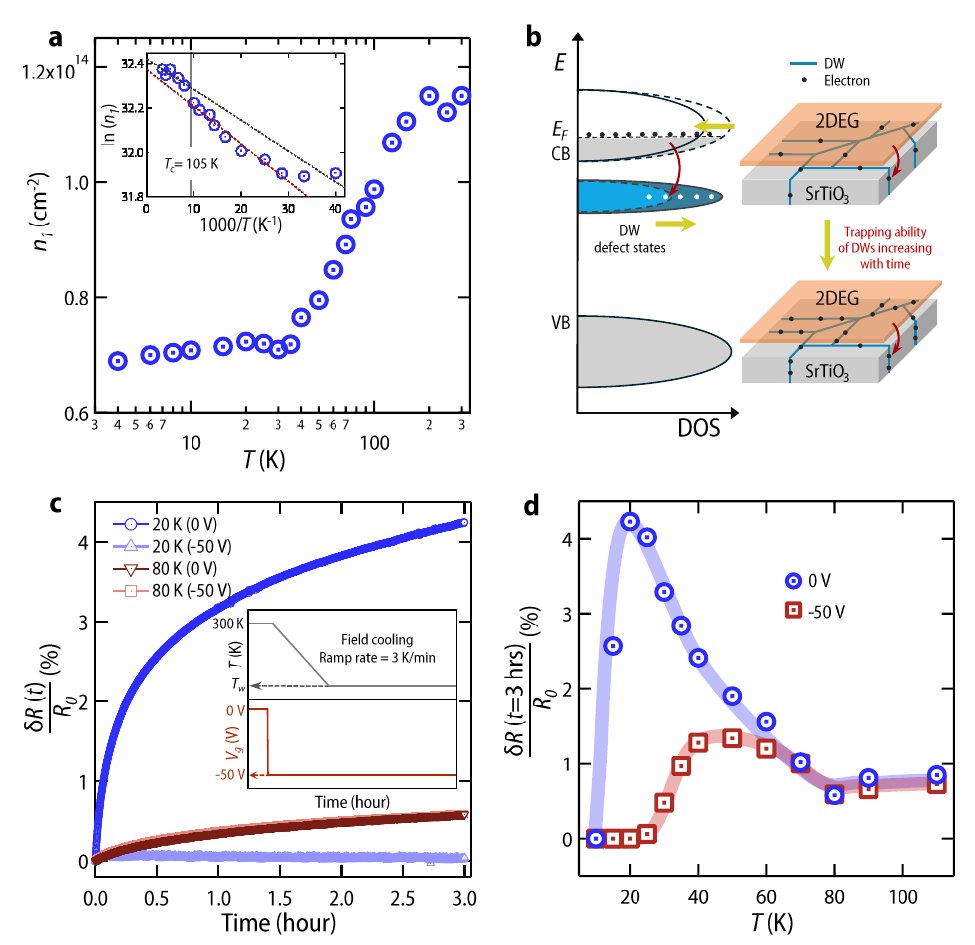}
			\caption{\textbf{Evidence of polarity} \textbf{a.} Sheet carrier density ($n_1$) of the dominant carrier density channel as a function of temperature showing carrier freezing below 150 K. We emphasize that our sample exhibits two-band transport below 90 K, and therefore, $n_1$ at 90 K and above was determined using one band model, whereas below 90 K it corresponds to the dominant carrier density channel obtained from the two band fitting. Inset shows Arrhenius plot of ln($n_1$) versus 1000/$T$ for the temperature range 300 K to 25 K ~\cite{Guduru:p2013p241301}. Black dotted line shows linear fit upto 125 K while red dotted line shows linear fit between 100 K and 35 K, where majority of trapping happens in domain wall defect states.  \textbf{b.} Schematic shows the proposed mechanism as the origin of slow relaxation of resistance. For a fixed phonon bath temperature, DW defect states increase with decreasing effective temperature of the DWs. As a result polarity of DWs increases and more and more electrons are getting trapped in DWs from 2DEG. The green arrow indicates that the DWs' DOS is increasing over time. \textbf{c.} Temporal evolution of resistance at fixed temperatures of 20 K and 80 K after zero field cooling and field cooling. The temperature ramping protocol for zero field cooling is similar to Fig. \ref{fig:2}b and for field cooling (shown in the inset), the system was cooled down from room temperature in the presence of a back gating voltage ($V_g$),- 50 V to a fixed temperature ($T_w$) ) at which the temporal evolution of resistance is measured. \textbf{d.}  The relative percentage change of resistance at the end of three hours $\frac{\delta R \textnormal{(t=3 hrs)}}{R_0}(\%)$ for several fixed temperatures for zero field cooling (blue) and field cooling (red).}\label{fig:4}}
	\end{figure*}

 {\color{magenta}{\bf{Relaxations of resistance:}}} 
 Following the observation of a static influence of DWs in the temperature-dependent resistance measurements, we now present evidence of their dynamical impact on conduction electrons. To investigate this, we cooled down the system at a ramp rate of 3 K/min from the room temperature to a desired low temperature (denoted as $T_w$) using a close-cycle cryostat and monitored the temporal evolution of resistance at the fixed final temperature (Fig. \ref{fig:2}a).
 The inset of the lower panel of Fig. \ref{fig:2}a shows one such recorded data where the $T_w$ was set to 20 K. As evident, the resistance relaxes gradually and doesn't saturate in laboratory time scale, mirroring the behavior seen in conventional electron glasses~\cite{Dobrosavljevic:2012p,Pollak:2013p,Amir:2011p235,RevModPhys.57.287}. To gain a deeper understanding, we performed identical set of measurements at various fixed temperatures encompassing the distinct phases of STO DWs and all the way down to 5 K. Notably, after each measurement, the sample was warmed to 300 K to ensure the recovery of its initial resistance value. To quantitatively capture the temperature evolution of these relaxations, we plot the relative percentage change of resistance at the end of three hours $\frac{\delta R \textnormal{(t=3 hrs)}}{R_0}(\%)$ in the upper panel of Fig. \ref{fig:2}b. As evident, with decreasing temperature, $\frac{\delta R \textnormal{(t=3 hrs)}}{R_0}(\%)$ remains almost temperature independent roughly up to 80 K and exhibits a non-monotonic enhancement with a maximum occurring around 22 K. A similar behavior was further confirmed for a 5 uc GAO/STO sample and another 15 uc GAO/STO sample (discussed in Fig. \ref{fig:4}). As can be further seen in the upper panel of Fig. \ref{fig:2}b, all the relaxations cease to exist below 12 K. To quantify the rate of relaxation as a function of $T$, we have also extracted $S$ =$ \frac{d\left[\frac{\delta R(t)}{R_0}\right]}{d\ln(t)}$ in the logarithmic regime of the relaxation.  $S$ exhibits similar temperature dependence as $\frac{\delta R \textnormal{(t=3 hrs)}}{R_0}(\%)$ (upper panel of Fig. \ref{fig:2}b).

As the compressor was on during all of these measurements, the mechanical vibration of the cryostat is sufficient enough to depin STO domain walls, as discussed by Kustov et al. in Refs. ~\onlinecite{Kustov:2020p016801,Salje:2023p1588}.
In a remarkable coincidence, the distinctive features in the temperature-dependent $\frac{\delta R \textnormal{(t=3 hrs)}}{R_0}(\%)$ and $S$ exhibit nearly a one-to-one correspondence with the established phase diagram of DWs of insulating STO, reported in Ref. ~\onlinecite{Kustov:2020p016801}. In the lower panel of Fig. \ref{fig:2}b, we also plot the low-amplitude ultrasonic internal friction for pristine STO (taken from the reference~\cite{Kustov:2020p016801}) for easy comparison.  The internal friction  directly probes the different forces resisting the motion between DWs and  has identified four distinct phases~\cite{Pesquera:2018p235701,Kustov:2020p016801}, marked in the lower panel of Fig. \ref{fig:2}b.
 Among these phases, the emergence of the domain glass phase is intriguing, particularly considering that in this temperature range, quantum fluctuations significantly influence the properties of STO~\cite{Chandra:2017p112502}, an entity that is generally considered an inhibitor of glass formation~\cite{Pastor:1999p4642}.

We now discuss how these collective domain wall phases manifest in transport measurements. Distinguishing between domain liquid and pinned phases is currently difficult. However, the increased resistance relaxation observed below 40 K, with complete suppression below 12 K, roughly coincides with the transition to the quantum domain glass and quantum domain solid phase, respectively.  
To explore glassy electron dynamics outside the standard electron glass framework~\cite{Davies:1982p758}, the remainder part of this paper concentrates on the domain glass phase.

  \begin{figure*}[htp]
		\centering{
			{~}\hspace*{-0.60cm}
			\includegraphics[scale=.75 ]{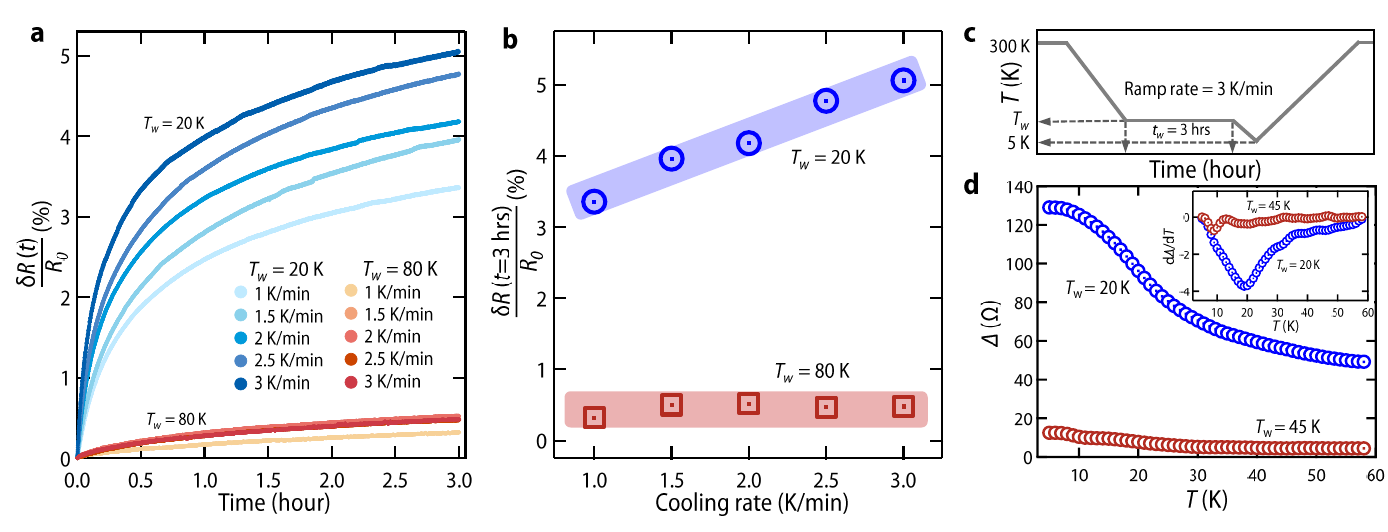}
			\caption{\textbf{Cooling rate dependence, memory effect, and evidence of glassy dynamics} \textbf{a.} Cooling rate dependent resistance relaxation observed for 3 hours at 20 K and 80 K. \textbf{b.} Relative percentage change of resistance at the end of three hours $\frac{\delta R \textnormal{(t=3 hrs)}}{R_0}(\%)$ for different cooling ramp rates. Shaded regions are included in the plot as guide to eye. Similar measurements for another 15 uc GAO/STO sample with cooling rate down to 0.1 K/min shows that $\frac{\delta R \textnormal{(t=3 hrs)}}{R_0}(\%)$ has a logarithmic dependence on the cooling rate. \textbf{c.} Temperature ramping protocol for studying the memory effect. Here the sample is cooled from room temperature to a waiting temperature $T_w$ (ramp rate 3 K/min) where the system is annealed for a time $t_w$=3 hours. Thereafter, it is cooled down to the base temperature of 5 K and heated back to room temperature \textbf{d.} Difference ($\Delta$) in resistance of the system obtained by subtracting the $R$ vs $T$ plot for $t_w$=0 (obtained in the heating run) from the one measured for $t_w$=3 hours. Inset shows the derivative of $\Delta$ with respect to $T$.} \label{fig:3}}
	\end{figure*}

{\color{magenta}{\bf{History dependence and confirmation of glassy dynamics:}}} 
Apart from the slow relaxation, another hallmark of
glassy dynamics is the dependence of their relaxation on the
previous history, the system has gone through ~\cite{Amir:2011p235,Dobrosavljevic:2012p,Pollak:2013p}. The
first evidence of this for the present study comes from our
cooling ramp rate-dependent measurements wherein we carried
out measurements similar to those shown in Fig. \ref{fig:2}a  with several cooling ramp rates ranging from 1 K/min to 3 K/min. Fig. \ref{fig:3}a shows the two sets of measurements for two fixed final temperatures $T_w$= 20 K and 80 K. As evident, resistance relaxation becomes slower with decreasing ramp rate for $T_w$ = 20 K (which is in the domain glass phase). On the contrary,  no appreciable changes are observed for $T_w$ = 80 K, which is outside the glassy phase. Quantitatively, this is even more apparent in Fig. \ref{fig:3}b, where we have plotted the relative percentage change of resistance at the end of three hours of relaxation for different cooling ramp rates.
Cooling ramp rate-dependent relaxation at $T_w$ = 20 K has also been measured for another 15 uc GAO/STO sample with a cooling rate down to 0.1 K/min. A strong ramp rate-dependence has also been observed there.
Additionally, we observe that the $R_0$ also increases systematically with the decreasing cooling rate. Cooling ramp rate-dependent relaxation at $T_w$ = 20 K has also been verified for the 5 uc GAO/STO sample.

Another systematic approach to confirm glassy dynamics is studying the memory effect~\cite{Amir:2011p235}. To investigate this, we have followed a temperature ramping protocol as depicted in Fig. \ref{fig:3}c. Here, we cooled down the 15 uc GAO/STO sample from the room temperature to a waiting temperature $T_w$ ($<$ 300 K) and waited there for the next 3 hours. Thereafter, we cooled down further to the base temperature (5 K) and recorded the resistance in the heating run. To capture the effect of waiting, we subtracted the measurement performed without any waiting from the waiting one. The difference ($\Delta$) of these curves has been plotted in Fig. \ref{fig:3}d for two values of $T_w$. We further emphasize that 20 K lies well within the domain glass phase whereas 45 K is outside this phase. It is clearly evident that $\Delta$ shows strong temperature dependency for $T_w$ = 20 K, which is negligible  for the measurement with $T_w$ = 45 K. Interestingly, the temperature derivative of $\Delta$ for $T_w$ = 20 K shows a dip at the same temperature where annealing was performed. However, no such dip was observed for $T_w$ = 45 K. These observations affirm the retention of memory effect of electrical resistance within the domain glass phase of STO.

 {\color{magenta}{\bf{Discussions:}}}
 Conventional electron glass models, rooted in the competition between electron correlation and disorder, fail to explain the glassy electron behaviour observed in metallic GAO/STO samples. Therefore, we propose a new mechanism centred on the polar domain walls (DWs) of STO as the distinct DW phases of pristine STO are intrinsically tied to the development of polarity within DWs~\cite{Pesquera:2018p235701, Kustov:2020p016801}. Moreover, these polar DWs effectively function as defect states by trapping charges~\cite{Nataf:2020p634}.
Since the domain wall polarity increases with the lowering of temperature~\cite{Scott:2012p187601,Salje:2013p247603,Frenkel:2017p1203},  more and more trapping of free carriers can be anticipated, leading to a reduction in free carrier density. To testify the same, we have measured Hall resistance at different fixed temperatures  (see Ref.~\cite{Mandal:2021p12968} for the method of Hall analysis). The temperature-dependent sheet carrier density of the dominant carrier density channel ($n_1$) indeed reveals a carrier freezing effect below 150 K (Fig. \ref{fig:4}a). We note that the carrier freezing starts at a little higher temperature than the $T_\textnormal{C}$ of STO. This can be ascribed to the trapping of thermally excited carriers in additional defect states created by oxygen vacancy clusters which onset at higher temperatures~\cite{Yin:2020p017702,ojha:2021p054008}. Nonetheless, the majority of trapping happens below 100 K [Inset of Fig. ~\ref{fig:4}a], indicating the presence of polar domain walls in our metallic GAO/STO samples.

 Before delving further into the possible mechanism, certain aspects of pristine STO are crucial to take into consideration. The presence of electrical dipoles within the highly polarizable STO very often leads to the formation of polar nano region (PNR) spanning several unit cells~\cite{Chandra:2017p112502}. 
 The resulting system exhibits a complex phase diagram, transitioning from a dipolar glass at low concentrations of PNRs to long-range ferroelectric order with overlapping PNRs at higher concentrations~\cite{Vugmeister:1990p993,Samara:2003pR367}.
 The quantum domain glass phase of pristine STO, which appears around 40 K~\cite{Kustov:2020p016801}, is indeed akin to such a dipolar glass phase~\cite{Viehland:1991p71,Salje:2016p163}. Crucially, such dipolar glass phase can survive even in electron-doped quantum paraelectrics~\cite{Ojha:2024p3830} and hence would be crucial in understanding the concomitant origin of glassy electron dynamics in the present case as discussed below.

One of the basic attributes of a glassy system is that once it’s cooled down to the glassy phase, the effective temperature of the system always remains higher due to its inability to equilibrate with its surrounding phonon bath~\cite{Leuzzi:2007thermodynamics,Mauro:2009p75,Leuzzi:2009p686}. Extending this analogy to our system, the domain wall in the domain glass phase can be considered to have an elevated effective temperature. Over time, it seeks to reach thermal equilibrium with the cryostat's set point.  As the domain wall polarity increases with decreasing temperature, this equilibration process would manifest as a gradual, glass-like growth in domain wall polarization with time. As the domain wall trapping ability is also strongly linked to its polarity, this temporal evolution consequently leads to a slow, glass-like increase in free carrier trapping, leading to the observed resistance relaxation in our experiments (see Fig. \ref{fig:4}b for the proposed mechanism). Notably, the resistance relaxation ceases to exist below 12 K, aligning with the quantum domain solid phase in bulk STO.

To confirm that the observed glassy electron dynamics originate from the background of the dipolar glass phase, we have further attempted to tune electron dynamics by applying an external electric field. It is well established that cooling a system of randomly oriented dipoles, such as in a dipole glass, in a biasing field can suppress the influence of the random fields~\cite{Samara:2003pR367}, thereby eliminating the glassy behavior. This occurs due to the alignment of the dipoles and the increase in their correlation length, potentially leading to long-range polar order. To investigate this, we used another 15 uc GAO/STO sample, which shows similar static and dynamic impact of domain wall in transport measurement. Fig. \ref{fig:4}c shows relaxation of resistance at fixed temperatures 20 K and 80 K after zero-field cooling (ZFC) and field cooling (FC) with a back gate voltage $V_g$= - 50 V from 300 K. The electric field was applied using conventional back-gate geometry~\cite{ojha:2021p054008}. At $T_w$ = 20 K (well within the domain glass phase), a significant reduction in resistance relaxation is observed upon FC compared to ZFC. However, at 80 K (far away from the domain glass phase), no difference is observed between FC and ZFC measurements. Fig. \ref{fig:4}d quantifies the relative percentage change in resistance after 3 hours for both ZFC and FC protocols at various fixed temperatures. As evident, the electric field has negligible influence on resistance relaxation above 60 K. However, below 40 K, FC drastically suppresses resistance relaxation compared to ZFC. This strongly emphasizes the definite role of the background dipolar glass in the observation of glassy electron dynamics, and we coin this newly identified electronic phase as dipolar glassy metal.

\begin{figure}
		\centering{
			{~}\hspace*{-0.2cm}
			\includegraphics[scale=.85 ]{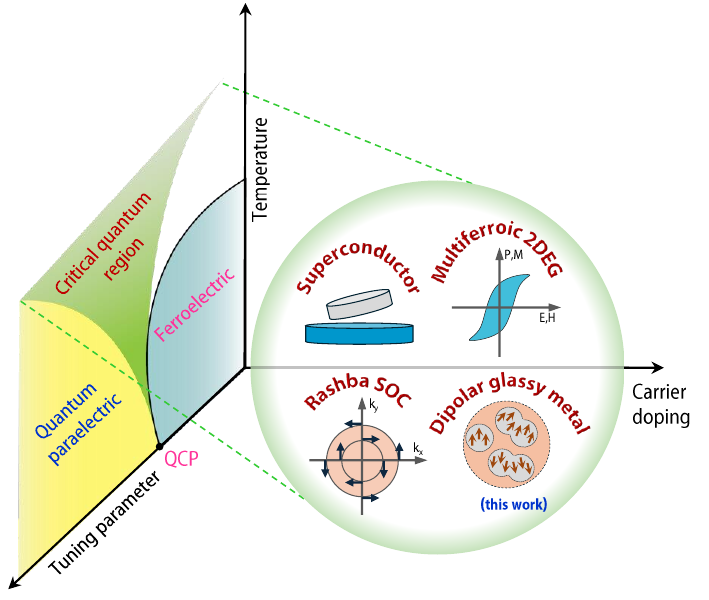}
			\caption{{\bf Phase diagram.} The well-known phase diagram with ferroelectric, quantum paraelectric and quantum critical region~\cite{Rowley:2014p367} together with an additional axis of carrier doping. This depicts a wide range of quantum phases that emerge  upon electron doping in SrTiO$_3$ through heterostructuring. Superconductivity, Rashba SOC and multiferroic 2DEG were reported in Refs.~\cite{Reyren:2007p1196,Caviglia:2010p126803,Brehin:2023p823} (also see Ref.~\cite{Pai:2018p036503}).This work presents the first evidence of a dipolar glassy metal phase in STO based 2DEG.} \label{fig:5}}
		
	\end{figure}

 {\color{magenta}{\bf{Conclusions and outlook:}}}
  In summary, we have demonstrated the first evidence of glassy electron dynamics in two dimensions arising from the complex interactions among polar ferroelastic twin walls of STO. Our finding of this dipolar glassy metal phase is in sharp contrast to the conventional electron glass phenomenology, where the glassiness emerges from the competition between random disorder and Coulomb interaction and vanishes well before the insulator-metal transition. This study not only broadens the spectrum of distinct quantum phases exhibited by STO-based conducting interfaces (Fig. \ref{fig:5}) but also underscores the unique aspects of electron transport in systems approaching a ferroelectric quantum critical point~\cite{Rowley:2014p367}. Furthermore, our observations will be pivotal in advancing our understanding of electronic transport in polar metals and superconductors~\cite{Bhowal:2023p53}. Very recently, topological defects in structural glasses have been found to explain some of the peculiarities of these systems~\cite{Baggioli:2023p2956, Baggioli:2021p015501}. In that context, since ferroelastic twin walls are naturally occurring topological defects~\cite{Nataf:2020p634}, our work opens up an avenue to explore the role of topological defects in the electron glass sector, which had remained completely unexplored to date.

	\noindent\section*{Methods}
	\noindent\textbf{Sample preparation}: All GAO films have been epitaxially grown on a mixed-terminated single-crystalline STO (001) substrate $(5 \times 5 \times 0.5\,\text{mm}^3$) (purchased from Shinkosha Co., ltd, Japan) utilizing a pulsed-laser deposition system~\cite{ojha:2021p054008}.  The sample was grown at a substrate temperature of $500^\circ\text{C}$ under a vacuum of approximately $10^{-6}$ Torr. A KrF excimer laser ($\lambda$=248 nm) was used for ablating the single crystalline GAO target, purchased from Shinkosha Co., ltd, Japan. The crystallinity of the film was checked using a lab-based Rigaku Smartlab X-ray diffractometer. X-ray reflectivity measurement was employed to determine the thickness of the film. 
 
	\noindent\textbf{Transport measurement}: Hall bar structure was made along the [100] crystallographic direction by selectively scratching the film all the way down into the substrate~\cite{Ojha:2023p126}. All the temperature-dependent electrical transport measurements related to resistance relaxation were carried out in a close cycle cryostat (Advanced Research Systems, USA) using a dc delta mode with a Keithley 6221 current source and a Keithley 2182A nanovoltmeter. A Keithley 2002 multimeter was used for measuring two probe resistance of the 1.5 uc and 1.75 uc films.  Magnetic field-dependent Hall measurements were performed using an Oxford Integra LLD system. For gate field-dependent measurements, an 80 nm Au layer was sputtered as a back electrode on the bottom side of the STO substrate. A Keithley 2450 source meter was used for applying the back gating voltage.

\noindent\textbf{HAXPES measurement}:  HAXPES experiments were carried out using a high-resolution Phoibos electron
analyzer~\cite{Schlueter:2019p040010}. The measurement chamber pressure was $\sim$ 5$\times$10$^{-10}$ Torr and the sample temperature was maintained at 40 K using an open cycle helium flow cryostat. The measured kinetic energy was corrected using Au 4$f$ core level spectra, acquired from a gold reference sample mounted alongside the GAO/STO heterostructure. For 3.4 keV photon energy, the inelastic mean free path ($\lambda$) of the photoelectrons is estimated to be in the range of 5-6 nm~\cite{Tanuma:2011p689}. This allows to access  Ti oxidation states near the interface for the 6.8 nm GAO/STO film as the effective probing depth is roughly 3$\lambda$.

	\section*{Data availability}
	The data that support the findings of this work are available from the corresponding authors upon reasonable request.
	
	\section*{Acknowledgements}
We thank Prof. Sumilan Banerjee for insightful discussions. The authors acknowledge the use of central facilities of the Department of Physics, IISc, funded through the FIST program of the Department of Science and Technology (DST), Gov. of India and the wire bonding facility of MEMS packaging lab, CENSE, IISc. JM acknowledges UGC, India for fellowship. MB and NB acknowledge funding from the Prime Minister’s Research Fellowship (PMRF), MoE, Government of India. SM acknowledges funding support from a SERB Core Research grant (Grant No. CRG/2022/001906) and I.R.H.P.A Grant No. IPA/2020/000034. The superconducting magnet, used for the magneto transport measurement was procured through a DST Nanomission grant (DST/NM/NS/2018/246) to SM. Portions of this research were carried out at the light source PETRA III DESY, a member of the Helmholtz Association (HGF). Financial support by the Department of Science \& Technology (Government of India) provided within the framework of the India@DESY collaboration is gratefully acknowledged.

	\bibliographystyle{naturemag}
	\bibliography{STOref_Glass}

\end{document}